\documentclass[conference]{IEEEtran}
\usepackage{ifpdf}
\usepackage{cite}
\ifCLASSINFOpdf
  \usepackage[pdftex]{graphicx}
  \graphicspath{{Figures/}}
  \DeclareGraphicsExtensions{.pdf,.jpeg,.png}
\else
\fi

\usepackage{graphicx}
\usepackage{tabularx}
\usepackage{booktabs}

\usepackage{array}
\begin{document}
\title{ Responsibility and Tangible Security: Towards a Theory of User Acceptance of Security Tokens}

% \author{
% \IEEEauthorblockN{Jeunese Payne}
% \IEEEauthorblockA{University of Cambridge\\
% jp662@cam.ac.uk}
% \and
% \IEEEauthorblockN{Graeme Jenkinson}
% \IEEEauthorblockA{University of Cambridge\\
% gcj21@cam.ac.uk}
% \and
% \IEEEauthorblockN{Frank Stajano}
% \IEEEauthorblockA{University of Cambridge\\
% fms27@cam.ac.uk}
% Õ\andÕ
% \IEEEauthorblockN{Angela Sasse}
% \IEEEauthorblockA{University College London\\
% a.sasse@cs.ucl.ac.uk}
% \and
% \IEEEauthorblockN{Max Spencer}
% \IEEEauthorblockA{University of Cambridge\\
% ms955@cam.ac.uk}
% }


\author{\IEEEauthorblockN{Jeunese Payne\IEEEauthorrefmark{1},
Graeme Jenkinson\IEEEauthorrefmark{1},
Frank Stajano\IEEEauthorrefmark{1}, 
M. Angela Sasse\IEEEauthorrefmark{2} and
Max Spencer\IEEEauthorrefmark{1}}
\IEEEauthorblockA{\IEEEauthorrefmark{1}University of Cambridge\\
Cambridge, United Kingdom
\\ Email: firstname.lastname@cl.cam.ac.uk}
\IEEEauthorblockA{\IEEEauthorrefmark{2}University College London, London, United Kingdom\\
Email: a.sasse@cs.ucl.ac.uk}
%Telephone: (800) 555--1212, Fax: (888) 555--1212}
%\IEEEauthorblockA{\IEEEauthorrefmark{4}Tyrell Inc., 123 Replicant Street, Los Angeles, California 90210--4321}
}


\IEEEoverridecommandlockouts
\makeatletter\def\@IEEEpubidpullup{9\baselineskip}\makeatother
\IEEEpubid{\parbox{\columnwidth}{Permission to freely reproduce all or part
    of this paper for noncommercial purposes is granted provided that
    copies bear this notice and the full citation on the first
    page. Reproduction for commercial purposes is strictly prohibited
    without the prior written consent of the Internet Society, the
    first-named author (for reproduction of an entire paper only), and
    the author's employer if the paper was prepared within the scope
    of employment.  \\
    USEC '16, 21 February 2016, San Diego, CA, USA\\
    Copyright 2016 Internet Society, ISBN 1-891562-42-8\\
    http://dx.doi.org/10.14722/usec.2016.23003
}
\hspace{\columnsep}\makebox[\columnwidth]{}}

\maketitle

\begin{abstract}
%\boldmath
Security and usability issues with passwords suggest a need for a new authentication scheme. Several alternatives involve a physical device or token. We investigate one such alternative, Pico: an authentication scheme that utilizes multiple wearable devices.  We present the grounded theory results of a series of semi-structured interviews for exploring perceptions of this scheme.
We found that the idea of carrying physical devices increases perceived personal
responsibility for secure authentication, making the risks and inconvenience
associated with loss and theft salient for participants. Although our work is
focused on Pico, the results of the study contribute to a broader understanding
of user perception and concerns of responsibility for any token-based authentication
schemes.
%Security designers wishing to challenge the status quo of passwords need to consider how to reduce the anxiety surrounding responsibility and highlight the benefits of tangible, token-based authentication schemes.
\end{abstract}


% \fi
%%\IEEEpeerreviewmaketitle

\section{Introduction}
Most users gain access to services using text-based passwords. This is despite wide recognition that passwords are impractical from a usability standpoint and vulnerable to attack from a security standpoint. Research by Adams and Sasse \cite{adams1999users}, Beautement \textit{et al.} \cite{beautement2009compliance}, and Weirich and Sasse \cite{weirich2001pretty} suggests that users hold inaccurate notions of security and its importance, and are thus less likely to adhere strictly to security policies. Users also often struggle to use different passwords for different accounts, each sufficiently long with a combination of letters, numbers, and symbols \cite{stajano2011pico}.

Even strong passwords are routinely compromised through malware and phishing attacks. Thus, rather than longer and more complex passwords, we require a stronger authentication scheme. We evaluate an alternative and more secure token-based system designed to eliminate passwords: \emph{Pico} \cite{stajano2011pico}. It has two key elements. First, the user scans a QR code that appears in a browser for web authentication with a small, portable, dedicated \textbf{authentication token} called \emph{Pico}. This token should have a display, button, and camera for taking a picture of the QR code. Second, the Pico is protected from the consequences of loss or theft by smaller, wearable \textbf{devices} called \emph{Picosiblings}; the Pico automatically locks when not in close proximity to \emph{K out of N} of these devices.

% Bonneau \textit{et al.}'s \cite{bonneau2012quest} comparative evaluation of various schemes scores Pico highly in security relative to passwords, but low on usability. 

It is difficult to encourage users and service providers to consider adopting Pico because we are not only challenging passwords as they are \emph{supposed} to be used, but as they are \emph{actually} used, which is more usable, though less secure, than Pico. The first task is to identify \textit{initial} reactions to a new scheme such as Pico, before dealing with long-term use. Our research aims to explore the acceptable design space of Pico prior to implementation, using low-fidelity prototypes to provoke discussion about its perceived usability and security.

To do this, we examined data from semi-structured interviews using grounded theory (GT), first developed by sociologists Glaser and Strauss \cite{glaser1967} and developed further by Strauss and Corbin \cite{strauss1998basics}. We used GT because it allows for systematic and empirical theory-creation \cite{corbin1990grounded}\cite{rich2012inside}, making it suited to exploratory research such as this. GT does not extend or test an existing theory; it is used to \emph{formulate} a theory based inductively on participant concerns and how they try to resolve them. Thus, despite the use of prototypes, the aim was not to involve participants in the design of tokens, but to gain insights into their opinions and concerns about a token-based system such as Pico. 

%These concerns were often not specific to either prototype of the Pico device, but reflected general concerns relating to usability and security.
% Since we did not know what participants would focus on, this research was primarily exploratory in nature.

In adopting this methodology, our research makes three contributions. First, while GT has already been adopted in examining password practices by such researchers as Adams \emph{et al.} \cite{adams1999users} \cite{adams1997making} and Stobert and Biddle \cite{stobert2014password}, we extend this to token-based authentication; in doing so, we contribute to the field by providing insight into factors influencing the acceptability of token-based schemes. Second, we not only reveal \textit{what} opinions emerge, but \textit{why} they emerge. Finally, we present a final GT model that systematically analyzes participant perceptions of Pico in terms of personal responsibility and the tangibility of security using tokens. To our knowledge, the issue of personal responsibility has not been formally identified in literature with regards to security tokens. 

\section{Background}

The number of passwords users are asked to create and maintain pushes beyond what people can reasonably be expected to remember \cite{adams1999users}, especially given the inconsistency between password policies \cite{florencio2010security}\cite{shay2010encountering}. As research by Adams \textit{et al.} \cite{adams1999users}\cite{adams1997making}, Shay \textit{et al.} \cite{shay2012correct}, and Stobert and Biddle \cite{stobert2014password} indicates, users rely on coping strategies, such as reusing and storing passwords. Even passwords created within password policy guidelines can be easy to crack using dictionary and brute-force attacks because users fulfill requirements in predictable ways \cite{burr2004electronic}\cite{shay2012correct}. To tackle these limitations, alternative types of passwords have been proposed. 

A pass\emph{phrase} -- a sequence of words -- is typically longer, harder to guess, and easier to remember than a traditional password \cite{keith2007usability}\cite{shay2012correct}. However, Shay \textit{et al.}\ \cite{shay2012correct} found that assigned passphrases did not outperform assigned passwords on usability, and users tended to write them down. Additionally, Keith \textit{et al.} \cite{keith2007usability} found that user-generated passphrases and passwords resulted in similar levels of recall failure, and that passphrases induced more typographical errors and were perceived as being harder to use than passwords. Additionally, similar to traditional passwords, passphrases are harder to keep track of and remember as the number of accounts increases, assuming the user tries to create different passphrases for different accounts. 

Some argue that \emph{graphical} passwords are easier to manage and to keep secure than text-based passwords. The argument is that pictures are remembered better and for longer than words, making graphical passwords more manageable than text-based passwords \cite{de2005picture}\cite{bonneau2012quest}. However, visual memory capacities may be over-estimated and, like text-based passwords, graphical passwords are often highly predictable \cite{de2005picture}, as well as less accessible for the visually impaired \cite{bonneau2012quest}. 

%Another attractive alternative is biometrics -- physiological and behavioral characteristics \cite{de2005picture}\cite{bonneau2012quest}. Using fingerprint recognition as an example, Bonneau \textit{et al.}\ \cite{bonneau2012quest} describe how the user need not consciously remember anything, and how the number of accounts does not increase the burden of authentication. However, biometric schemes do not score well on other security dimensions such as resistance to theft and phishing, and also evoke privacy concerns \cite{de2005picture}\cite{bonneau2012quest}.

Instead of passphrases and graphical passwords, perhaps all that is needed is a
usable method of storing unique and secure alphanumeric passwords, such as a
portable password manager. This has drawbacks, including requiring the user to
transcribe text and to carry an extra device. Phone-based password managers
might be more convenient to carry, and, according to Karole \textit{et al.}
\cite{karole2011comparative}, also increase trust because a phone is locally
controlled by the user. However, it still involves transcribing passwords into
the terminal. USB password managers are directly inserted into the terminal, but
a USB drive is still an extra device to carry, and not every device (such as a
tablet) has an available USB port. Although more usable, users in Karole
\textit{et al.}'s\ \cite{karole2011comparative} study did not like the idea of
using an online password manager because it requires trust in a third-party service provider that stores encrypted passwords on a remote server.

Recently, FIDO (Fast IDentity Online), an open industry alliance of vendors,
including Microsoft, PayPal, and Visa, released two sets of specifications for
online authentication: UAF (Universal Authentication Framework) and U2F
(Universal Second Factor). Both involve authenticating to online services with a
device. UAF replaces passwords entirely by allowing users to authenticate from a
FIDO-enabled device, such as a smartphone. It involves registering the user's
device to online services and selecting a biometric authentication action, such
as swiping a finger, performed on the device. U2F replaces the second factor in
two-factor authentication (2FA) with a dongle that the user inserts into a USB port. The user presses a button on that dongle to authenticate to services after typing a password.

Users tend not to like 2FA when it involves carrying a dedicated device. De Cristofaro \textit{et al.}\ \cite{de2013comparative} found that dedicated tokens were the least used method of 2FA: almost 90\% of participants used email or SMS; approximately 45\% used a phone app; less than a quarter used tokens. Most used a hardware token only because their employer or bank forced them to. This is in line with Krol \textit{et al.} \cite{krol2015they}, who found that carrying and operating a hardware token for 2FA with online banking negatively correlated with satisfaction. 

There is little research into user experiences and perceptions of entirely token-based authentication, despite there being a number of potential benefits over passwords. Most notably, these are security benefits: physical tokens tend to be resilient to physical observation and guessing \cite{bonneau2012quest}. They also alleviate memory issues associated with having to remember passwords and linking the right ones to the right accounts. There are, however, important drawbacks. Tokens require users to carry an additional item -- a usability issue \cite{bonneau2012quest}. There is also the potential for loss and theft. This is both an inconvenience (in terms of token-recovery) and a security challenge \cite{bonneau2012quest}\cite{karole2011comparative}: if physical tokens are not always something that users really ``have", then unauthorized persons could access services that belong to someone else. Picosiblings offer a potential solution to this, but also present a notable usability drawback: users are asked to carry \emph{multiple} devices to safeguard them against loss and theft. Is there scope to create a usable authentication scheme that relies on multiple physical devices?

\section{The Study}
Our study was conducted to develop a theory that explains the acceptability of token-based authentication. This was defined within the design space of Pico, aided by, but not limited to, the low-fidelity prototypes described in the following section. In particular, we are interested in how Pico is understood and how any concerns are resolved in the minds of participants, as this will give us insight into their mental models of token usability and security, informing the final GT.

%help us determine whether there is scope to create a \emph{usable} token-based authentication scheme that people might use. Do achieve this, we sought to answer the following research questions:
%\begin{enumerate}
%\item What is the theory that explains the acceptable design space of token-based authentication?
%\item How is the design space of Pico understood?
%\item How are any concerns about Pico resolved in the minds of participants?
%\end{enumerate}
%We examined the acceptable design space of multiple devices using the following method.
\subsection{Method}
Twenty semi-structured interviews were conducted, lasting between fifteen and thirty-five minutes depending on how much the participant wanted to communicate. The interview involved two preference tasks, asking participants to identify which items they might be willing to carry: for Pico itself and for the Picosiblings that unlock it. Each task was followed by questions about the items chosen. The participant's main points were iterated back to them; the interviewer asked if these were correct and if the participant had anything to add. 

The data were examined using GT, characterized by constant comparison of data to identify themes and categories that would otherwise be overlooked \cite{corbin1990grounded}. The theory is sufficiently developed when no new categories emerge during this comparative process -- theoretical saturation \cite{glaser1967}\cite{brown2002exploring}. GT occurs over three overlapping phases. Adhering to Strauss and Corbin's \cite{strauss1998basics} conception of coding, these stages include: 
\begin{enumerate}
\item Open coding. Dividing data into segments and looking for commonalities that reflect categories. The aim is to \textbf{reduce} the data into a smaller set of themes that appear to describe the topic under investigation;
\item Axial (focused) coding. Grouping discrete codes into conceptual categories that reflect commonalities and describe \textbf{relationships} between codes;
\item Selective coding. Identifying the \textbf{process} by which categories relate to a core category by selectively analyzing code clusters to form the GT.
\end{enumerate}

A number of procedures helped ensure the credibility of the analysis. First, notes taken during interviews were compared with corresponding video-recordings. This reduced interviewer bias, helped the researchers acquire a deep understanding of the core concepts, and ensured the veracity of the data. Second, we began with trial open-coding of the first six interviews. A consistency check was conducted where the primary researcher coded the interviews on two separate occasions (double-coding) and drafted an initial coding frame. A second coder (``peer debriefer") identified key themes without this coding frame (blind-coding)\footnotemark. Coding proper restarted from interviews one to sixteen. The primary researcher and peer debriefer made separate memos and came together to discuss the developing coding frame. Third, axial coding was expanded to include data from an additional four interviews, helping us determine whether the emerging theory adequately described the data and that theoretical saturation was reached. Fourth, the researchers consistently referred back to transcriptions, checking throughout the coding process that categories did emerge from the data. Finally, the primary researcher and peer debriefer sought diverse feedback on interpretations of the data and evolving theory from two ``inquiry auditors": another research team member and a colleague at a different university.

\footnotetext{There is typically no numerical inter-rater reliability reported with GT because codes are transitory and dynamic.} 

\subsection{Prototypes}
Two Pico designs and thirty-six Picosibling prototypes were created for the preference tasks conducted during the interviews, making the idea of carrying devices more tangible and easier to discuss. These prototypes are described below.

\subsubsection{Pico}
\begin{figure}[h]
\centering
\includegraphics[width=0.35\textwidth]{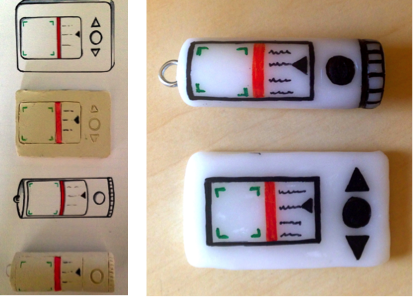}
\caption{Left: prototypes, drawn then modeled in plasticine. Right: prototypes modeled using Polymorph, used in the interviews.}
\label{Polymorph}
\end{figure}

Two prototypes were designed to be as comparable as possible in functions and features: same sized screen with green overlay for aiming at the QR code; LED strip and list of options; selection button; means of scrolling; and camera. These prototypes were drawn, then recreated in plasticine and Polymorph\footnotemark. Polymorph versions were used in the interviews (Figure~\ref{Polymorph}). 

\footnotetext{Small plastic granules that, when heated in hot water, form a transparent flexible material; this hardens again as it cools.} 

The main differences between these prototypes were in overall shape and scrolling function. The first Pico prototype was similar to a credit-card but 3mm thick to (theoretically) allow for a camera and screen in the hardware. This prototype included arrows for scrolling between accounts. The second prototype was cylindrical, roughly the length of the card-shaped prototype, with a scroll bar wrapped around one end. They could be easily re-oriented to suit participant handedness.

\subsubsection{Picosiblings}
\begin{figure} [h]
\centering
\includegraphics[width=0.35\textwidth]{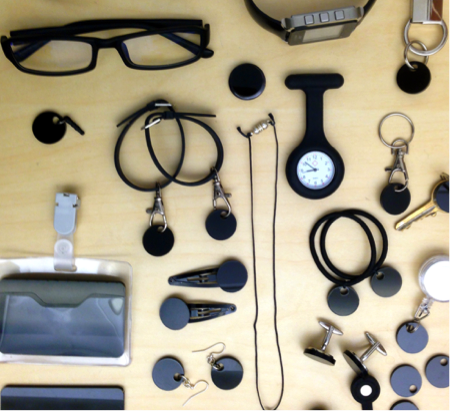}
\caption{Picosiblings used in interviews.}
\label{Plastic}
\end{figure}

There were thirty-six Picosibling items, such as glasses and watches, magnetic clips, small free-standing coin-shaped discs, and items to which some coin-shaped discs were attached, such as keys and hairbands. Most items were plain, made from black plastic, and shaped by a laser cutter for uniformity in design (Figure~\ref{Plastic}). Others were created or bought based on pilot participants' suggestions for items such as an ID card. 

\subsection{Procedure}
\begin{figure}
\centering
\includegraphics[width=0.45\textwidth]{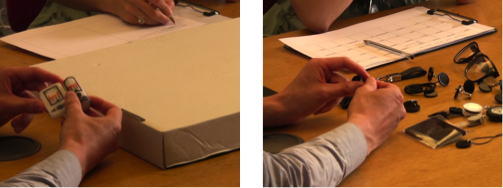}
\caption{Left: A participant comparing the two Pico prototypes; Right: A participant interacting with the Picosiblings.}
\label{Hands}
\end{figure}

Before beginning, participants gave informed written consent to partake in the study, which included having their hands video-recorded (Figure~\ref{Hands}), alongside their comments, as they interacted with the prototypes. Participants then answered short demographic questions and, at their own pace, navigated through a set of presentation slides introducing Pico and how it would work for logging into web accounts (typically less than 2 minutes). This avoided variability in the description of Pico. A video-recorder was then turned on and the semi-structured interview began.

Interviews consisted of nine core questions, each with candidate follow-up questions designed to encourage participants to give more information, such as ``why do you think that is?" and ``can you tell me more about...". Though the core questions followed on logically from each other, it was not necessary to ask the core questions strictly in that order: as is the nature of semi-structured interviews, if the participant took the discussion down a different route, the researcher could chose to ask a core question earlier or later than planned, so long as all core questions were eventually covered. Nevertheless, most interviews followed the structure listed below, and the first question was always the same\footnotemark:
\begin{itemize}
\setlength\itemsep{0.2em}
\item How participants found using and remembering passwords
\item Which of the two Pico prototypes they preferred
\item Whether carrying a standalone Pico was something the participant could imagine doing
\item The kinds of items they already carried with them
\item Why they chose particular Picosiblings
\item Whether there were other items they might have as Picosiblings
\item The number of Picosiblings they would be willing to carry with them at one time 
\item Any problems they envisaged with the system
\item Whether they had any suggestions for the design of our token-based scheme. 
\end{itemize}
Videos were checked against notes taken in each interview. The recordings were watched and listened to several times by more than one researcher, and were transcribed and coded to extract phenomena (events, objects, actions, ideas, issues, and experiences) of relevance to the potential use and design of our scheme. According to Glaser \cite{glaser1998doing}, and other researchers (e.g. \cite{kenealy2008management}\cite{halcomb2006verbatim}), there is no need to transcribe interviews in their entirety. Instead, a mixed-method of transcribing and analysis was employed -- ``selective transcription", where unrelated rapport-building conversation and interviewer questions themselves were not included \cite{halcomb2006verbatim}. 

\footnotetext{This preliminary question eased participants into the interview and into thinking about authentication.}

\subsection{Participants}
Participants were sampled from a population of people who use passwords to access online accounts, recruited using online advertisements and snowball sampling. To maximize the potential to discover as many concepts as possible, GT is often conducted using a highly variable sample \cite{strauss1998basics}\cite{brown2002exploring}. For this reason, in later stages of data collection and analysis, representative individuals were selected from particular sub-groups for their theoretical relevance -- their potential for introducing new data for developing categories. This is theoretical sampling, described by Glaser and Strauss \cite{glaser1967}, and Strauss and Corbin \cite{strauss1998basics}. Specifically, the aim was to maximize diversity in age (20-57), gender (10 male, 10 female), and occupation (Table~\ref{ppts})\footnotemark. In total, twenty participants were recruited; mean age was 30.5 years (29.2 for male participants; 31.8 for female). An Amazon voucher was offered to participants to thank them.

\footnotetext{Participant gender is included in Table I to give readers an idea of the variability of participants even when more than one had the same occupation. Participant ages are omitted to avoid participants being personally identifiable based on the aggregation of this information.} 

\medskip
\begin{table} [h]
	\centering
	\caption{Participant Counts By Occupation}

\begin{tabular}{ l c }
\toprule
\textbf{Participant Occupation (and gender)} & \textbf{Count}\\
\midrule
Accounting (female) & 1\\
Engineering (male) & 1\\
Military (male) & 1\\
Admin/Clerical (male) & 1\\
Publishing (female) & 1\\
Translating (female) & 1\\
Software Developer (male) & 1\\
Homemaker (female) & 1\\
Unemployed Software Engineer (male) & 1\\
Unemployed Product Designer (female) & 1\\
Researcher: Physics (male) & 1\\
Researcher: Neuronal Development (male) & 1\\
Researcher: Cancer (female) & 1\\
School Teaching Assistant (1 male, 1 female) & 2\\
Postgrad Student: Sustainable Energy (female) & 1\\
Postgrad Student: Comp Sci (1 male, 1 female) & 2\\
Student: undisclosed subject (male) & 1\\
No occupation given (female) & 1\\
\bottomrule
\end{tabular} 
\label{ppts}
\end{table}

\section{Results}
Open coding of the first six interviews (trial phase) involved selecting and labeling words and phrases to form codes and initial categories. The two coders were in agreement about the main categories and conducted open coding again, this time on interviews one to sixteen. Here we report the coding frame created during axial coding. Data collected from the final four interviews substantiated the core categories for the \emph{Pico Token}, and \emph{Picosiblings}, described here.

\subsection{Pico Token}
Responses related to the Pico token fell into three core categories: \emph{Convenience} of logging into services, \emph{Prototype Preferences} for a dedicated Pico, and \emph{Trustworthiness} in terms of reliability and security. 

\subsubsection{Convenience}

The main issues associated with the convenience of using Pico were: \emph{(i)} the \emph{Efficiency} of authentication; \emph{(ii) Deployment}, or the number and types of services and devices that could be used with Pico; and \emph{(iii)} Pico as \emph{Something to Carry}.

\medskip
\emph{i. Efficiency:} It was important that logging in with Pico would be low-effort and time-effective. In particular, participants expressed concern over the number of steps involved, dictated in part by having to interact directly with the Pico, which was perceived as awkward and unnecessary. Participant 19, for example, stated: 
\begin{quotation}
\noindent ``It should be very, very simple for people to use. And I didn't at all see the point of having to select things on [Pico]... I want to, like, maybe authorize and cancel." 
\end{quotation}
%In particular, there were two aspects of the current conception of Pico that appeared inefficient to participants: \emph{(a) Scanning a QR Code} and \emph{(b)} the requirement for \emph{Scrolling Between Accounts} on the token rather than the terminal. 

\medskip
\emph{ii. Deployment:} The acceptability of Pico was dependent on it being widely adopted. It was considered important that users could authenticate to most, if not all, services and devices before it was worth the effort of carrying and using Pico, captured in the following comment from Participant~5:
\begin{quotation}
\noindent ``In terms of investing in a piece of technology, it would need to be able to connect to everything that I use before I think it would be worthwhile investing in it"
\end{quotation}
Some suggested that they would use Pico only if they were forced to (e.g. for work) or because it was the only authentication method available.

One of the main issues was that participants did not see the relevance of Pico for mobile devices, since these typically provide users access to a number of services without requiring them to repeatedly login. This challenge for token-based authentication is summarized in the following comment from Participant~12:
\begin{quotation} 
\noindent``Most people these days are always on the move, and always logging in from things like their tablets or their phones... I can't see a real need to, like, bring [Pico]"
\end{quotation}

\emph{iii. Something to Carry:} The majority expressed a preference for a Pico that was integrated with some other device that they already carried. There were two main categories of suggestion: design Pico with a \emph{Dual-Purpose}, or create a \emph{Smartphone App}. Participants gave a number of suggestions for a Pico that was designed to be part of some other useful device, such as a watch or a debit/credit card, and to therefore serve some additional function other than authentication. For example, Participant~8 stated:
\begin{quotation}
\noindent ``I could see that it would be more use if it was built into something else... that you have to carry around with you anyway that just does something extra"
\end{quotation}
The most popular suggestion, however, was that Pico be integrated with their phone.

\subsubsection{Prototype Preferences}
While the majority of participants would prefer to carry a dual-purpose Pico or smartphone app, some described the idea of carrying a separate dedicated Pico to be not too much of a burden. The extent to which participants were comfortable with carrying a dedicated Pico was dependent on: \emph{(i) Familiarity} and \emph{(ii)} whether it would be \emph{Easy to Use, Hold, and Carry}.

\medskip
\emph{i. Familiarity:} User experience of other devices had an impact on their acceptance of a dedicated Pico, specifically: \emph{Familiarity of Concept} and \emph{Familiarity of Design}. Familiarity of concept refers to prior experience of carrying some form of security token, such as bank tokens. Familiarity of design refers to experience with physically similar devices. Generally, participants preferred the more familiar, card-shaped design to the more novel, cylindrical design because the card-shaped design was similar to the shape of other devices, such as MP3 players, phones, and cameras.

\medskip
\emph{ii. Easy to Carry, Hold and Use:} Preferences for one of the low-fidelity prototypes over the other -- typically the card-shaped prototype over the cylindrical prototype -- was also dependent on the physical effort associated with carrying, holding, and using it. This depended on: its \emph{Shape and Size}, and its \emph{Scrolling Method}. The most common comment was that the card-shaped prototype would fit well in a pocket, wallet, or purse, making it easier to carry. It was also suggested that it would be easier to hold and use because it was flat and had minimal mechanical parts. Though some liked the scroll bar on the cylindrical prototype, many were concerned about its breakability and about the ease with which it could be used relative to the buttons on the card-shaped prototype.

\subsubsection{Trustworthiness}
Users had high expectations of the \emph{(i) Reliability} and  \emph{(ii) Security} of Pico.

\medskip
\emph{i. Reliability:} Participants were concerned about scenarios in which Pico would not work, meaning they would potentially have no access to their accounts. One of the main concerns was whether Pico would consistently work as intended -- whether there would be connection issues or whether Pico would have a short battery lifetime. There were also concerns about physical durability, centered on whether Pico would be waterproof and could handle everyday physical strains.

\medskip
\emph{ii. Security:} Participants appeared to place particular emphasis on the security of Pico. Participants expressed concern about who controlled the data, and whether their data could be misused. Another source of anxiety, despite the proposed system of Picosiblings for locking and unlocking Pico, was impersonation of the user after loss or theft (see Section \ref{sec:routine-use} \emph{ii}).

Participant responses suggest that introducing some form of 2FA could lessen anxiety. A minority of participants suggested that the security risk could be minimized with biometrics, the implication being that biometric authentication methods were more secure. For example, Participant~6 asked:

\begin{quotation}
\noindent ``Don't they say that there is going to be iris recognition say for buying, to get your money out of the bank? Why can't you do that?"
\end{quotation}

\subsection{Picosibling Preferences}
Participants engaged in an analysis of the trade-off between the perceived benefit of Picosiblings, and the inconvenience of managing multiple items. Suggestions for minimizing the inconvenience formed the following three core categories: \emph{Hedonic Concerns, Utilitarian Concerns}, and \emph{Routine Use}.

\subsubsection{Hedonic Concerns}
Responses highlighted preferences for control over Picosibling appearance. Such responses implied that hedonic factors might increase the willingness of some users to adopt Pico by making additional devices \emph{desirable} in terms of the hedonic goals they might help users meet, and thus less of an inconvenience. Suggestions for achieving this fell into two subtly different categories: \emph{i. Self-Presentation} with wearable devices, and \emph{ii. Personalization}, with devices that can be changed by the user as they prefer. 

\medskip
\emph{i. Self-Presentation:} It was important that Picosiblings fit in with what participants already wore. Some suggested that they should come in a range of designs. These could then \emph{add to} a user's personal style and aid in their self-presentation; others suggested simple and discrete devices that \emph{did not interfere} with personal style. For example, one participant said they chose a necklace because there were times they might wear a dress and have few places to keep Picosiblings. In this example, the second function of the Picosibling was decorative.

\medskip
\emph{ii. Personalization:} The desire for personalization taps into the desire for novelty, fun, and creativity, improving user experience. It was suggested by one participant that personalizable Picosiblings might make good gifts. In this way, personalization could help users meet social goals, helping them express themselves by creating devices that fit in with their personal style and with the personal styles of others.

\subsubsection{Utilitarian Concerns}
Suggestions for mitigating the physical and cognitive effort associated with carrying Picosiblings in ways that met utilitarian needs fell into three categories: \emph{(i)} whether Picosiblings could serve a \emph{Dual-Purpose} (additional \textbf{results of use}, aside from security), \emph{(ii)} the \emph{Practical Convenience} of carrying particular devices (the \textbf{process of use}), and \emph{(iii)} the \emph{Flexibility} that these devices allow, referring to how easy they would be to transfer between items and replace when the user wants. The data reveal different levels of convenience achieved with each type of Picosibling (Table~\ref{Utilitarian}). 

\begin{table*}
	\centering
	\caption{Sources of convenience for different types of Picosiblings}
	\begin{tabularx}{\textwidth}{ p{0.125\textwidth} X X X }
		\toprule
		~ & Cognitive Effort & Physical Effort & Flexibility\\
		\midrule
		\raggedright Dual-Purpose (e.g. a watch) & Reduced: User carries items they would carry anyway. & Reduced: Devices are embedded within existing items. & Limited by the number of dual-purpose items the user has. \\
		\\
		\raggedright Practically Convenient (e.g. a keyring) & Reduced: User carries devices that can be carried with or attached to items they carry anyway. & Increased: Device is an additional item that the user makes space for to carry. &Increased in terms of replaceability; limited in the types of items that devices can be carried with.\\
		\\
		\raggedright Flexible (e.g. a sticker on a phone) & Increased: User must keep track of which items have a sticker or clip attached. & Reduced: Devices are integrated with existing items. & Increased in terms of replaceability, transferability, and location.\\
		\bottomrule
	\end{tabularx}
	\label{Utilitarian}
\end{table*}

\medskip
\emph{i. Dual-Purpose:} It was important to most participants that Picosiblings not only unlock their Pico, but help them meet other goals. A common suggestion was for Picosiblings embedded \emph{within} other items, such as watches, credit cards, or coins for trolleys -- to serve a dual-purpose. However, if such devices stopped working, were lost, or were forgotten, or if the user simply did not want to use the device any longer, these dual-purpose Picosiblings would not be as easy to replace as ``practically convenient" Picosiblings that \emph{attach to} existing items or more ``flexible" Picosiblings that are designed to stick or clip to items the user already owns, described below.

\medskip
\emph{ii. Practical Convenience:} The difference between this sub-category and the ``dual-purpose" sub-category is subtle: practical convenience is determined by whether the user can easily carry Picosiblings \textit{with} items they already carry; they need not be fully integrated into those items. Such items would be more convenient by imposing less on the user's life. However, this tended to be limited to a few items that people carry frequently, which included phone, wallet, and keys, and depended on physical attributes such as size, shape, and detectability. 

\medskip
\emph{iii. Flexibility:} Participant responses suggested that it would be important to allow potential users to move Picosiblings between various items. The main benefit of this flexibility would be allowing users more control over which items they would like to use as Picosiblings as opposed to relying on ready-made (dual-purpose) devices or finding (practically convenient) places to keep non-embedded devices. This relies on creating Picosiblings that are integratable and removable. A recurring suggestion was to create Picosiblings that could be easily stuck to and easily removed from belongings, an idea caught in the following proposal from Participant~8:
\begin{quotation}
\noindent ``If you could make it small enough to be like a sticker that you can put inside the back of your cover of your mobile phone... or in cars and stuff"
\end{quotation}

Similar suggestions included: Picosiblings inserted into and taken out of shoes, and a clip-in/clip-out scheme. However, more flexible devices would perhaps be harder to keep track of because their location could change frequently. 

\subsubsection{Routine Use}
\label{sec:routine-use}
Participants stressed the presumed effort associated with managing multiple devices. The majority of Picosiblings were considered in terms of how they would be routinely used \emph{(i)} on a \emph{Day-to-Day} basis, and \emph{(ii)} in terms of the \emph{Exceptions} that could interrupt this. These exceptions consistently focused on scenarios that related specifically to preventing or recovering from loss and theft. 

\medskip
\emph{i. Day-to-Day:} Participants expressed a preference for Picosiblings that they would be able to have and rely on all the time because they would be used daily, if not always. For example, Participant~7 suggested:
\begin{quotation}
\noindent ``If I can wear something that is permanent. Like, for me, the wedding ring is permanent"
\end{quotation}
The more permanent the device, the smaller the impact on cognitive load, captured in the following comment from Participant~17:
\begin{quotation}
\noindent ``And these [bracelets] would be a permanent fixture. You know that you've got them at all times. You wouldn't have to worry about them"
\end{quotation}

The items that users tended to carry with them most often were their keys, wallet, and phone. Thus, a recurring suggestion was that Picosiblings either belonged to or could be attached to at least one of these three items. Some Picosiblings were also chosen because they could be used frequently, if not everyday, such as during exercise (e.g. trainers or wristbands) or at work (e.g. cufflinks or lab glasses). 

Rather than relying solely on accessories and clothing, some suggested that Picosiblings might be kept in more fixed \emph{locations}, such as on a work desk, at home, or in a car. 

\medskip
\emph{ii. Exceptions (Loss and Theft):} Participants asked questions about loss and theft, and relatedly, forgetting Picosiblings; the main source of anxiety was with recovery.
 
Most expressed concern with the inconvenience of carrying Picosiblings, or at least did not see how the security benefit of such devices outweighed the cost. Many believed that this system was insufficient for protecting against impersonation. In particular, participants were concerned that a thief could simply steal their Picosiblings, along with their Pico. 

As well as particular items (notably, the magnetic clips), it was suggested that particular situations were more conducive to loss and theft. Participant~4, for example, commented:
\begin{quotation}
\noindent ``From working in a school, with teenagers especially, I can see that it would become a game for that age group to, you know,... hide them and access everyone's account"
\end{quotation}

Participants typically wanted a minimal number of Picosiblings, and did not want to manage too many spares. Other means of protecting the few Picosiblings participants would be willing to carry were proposed instead. The most common suggestion was for hidden or discrete Picosiblings. A less common suggestion was to have weighted (by importance) Picosiblings. For example, Participant~2 suggested: 
\begin{quotation}
\noindent``I don't know if it's worth, like, having a core one that you have to have... But you could choose that yourself so if, like, thieves got them, they wouldn't be like \emph{`ha! I got your main one'}"
\end{quotation}

% Weighted Picosiblings could protect, not only Pico from theft, but particularly important services. 
It was also suggested that user-defined hierarchies of important accounts should be respected in the number of Picosiblings required to unlock the Pico, captured in the following comment from Participant~8:
\begin{quotation}
\noindent ``So, some stuff maybe doesn't matter as much as other stuff. So,... bank accounts, or important email addresses or that kind of thing, you require more for. In that case I wouldn't mind having three separate [Picosiblings]... because I'd probably be accessing those from somewhere like home or from the office where I can keep one"
\end{quotation}

These comments echo the thinking behind having a different password for different services based on the perceived importance of these services. Believing that more devices means more security appeared to lead to a more natural mental model of this system, one where a hierarchy of accounts could be unlocked by the addition of more Picosiblings, with zero possibly being sufficient for some cases.

\subsection{The Emerging Theory}
Selective coding revealed an obvious core category (phenomenon) underlying impressions of Pico -- \textbf{Inconvenience}. This was intimately related to the effort that participants imagined they might have to expend on using and remembering to carry a dedicated Pico device. As such, a popular suggestion was to have Pico integrated with another device, such as a watch, or for it to be a smartphone app. Similarly, some participants implied that authentication devices should be kept with other access-centered physical items already carried by most people -- credit/debit cards and keys. For example:
\begin{quotation}
\noindent ``I like the card kind of idea, kind of because you can maybe put it [a card-shaped Pico] \textbf{with other secure... with your bank card}" (Participant 9)

\medskip
\noindent ``I guess the [Pico]... \textbf{has this kind of keyness to it} so it kind of makes sense [to keep Pico with your keys]" (Participant 13)
\end{quotation}

More devices -- Picosiblings -- increased perceived inconvenience because of the cognitive effort (keeping track of Picosiblings) and the physical effort (carrying Picosiblings) associated with managing them. For many, having additional Picosiblings that were easy to attach to existing items, or that had additional functions mitigated this trade-off. Also important was the flexibility afforded to potential users in terms of the transferability and replaceability of Picosiblings, though this might be less convenient in terms of the cognitive effort needed to keep track of them. 

%In general, the inconvenience associated with having to carry items in the first place was not in line with most participants' perceptions of risk to their accounts based on their current behaviour with passwords.

In considering the effort of managing physical devices, participants expressed concerns about the reliability and security of Pico, and the loss, theft, or forgetting of Pico and Picosiblings. Participants anticipated not only annoyance due to inconvenience, but anxiety due to perceived risk. These observations form the basis of the second phenomenon identified in selective coding: \textbf{Risk Perception}. To mitigate perceptions of risk to security of Pico, some participants made suggestions designed to provide additional reassurance. For example:
\begin{quotation}
\noindent ``Is there a way to do, like, a time thing on them? ... I guess it just \textbf{makes it even more secure} ... because it's \textbf{changing all the time}" (Participant 2)
\end{quotation}

\begin{quotation}
\noindent ``[Fingerprint recognition] will probably be a \textbf{physical connection} to that device... And it will have a \textbf{green light or red light}" (Participant 13)
\end{quotation}

%?Is there a way to do, like, a time thing on them? I know I have this thing at work that I have to use when I log into the system. It has a number that changes every 10 seconds. I guess it just makes it even more secure? because it?s changing all the time?; ?That?s a combination of like a number that changes and then my own password?; ?I can kind of understand that extra security step [with two factor authentication] and I?ve got that enabled? I have that for a reason...?; ?[Biometrics] will probably be a physical connection to that device? And it will have a green light or red light?; ?It seems pretty secure. You have to the siblings before it will work?; ?Don't they say that there is going to be iris recognition say for buying, to get your money out of the bank? Why can?t you do that??

These suggestions highlight a disconnect between user mental models of the security of Pico and Pico's \textit{actual} security. Similarly, with Picosiblings, the risk of losing or forgetting Picosiblings evoked anxiety, even though the \textit{K out of N} scheme should mean that users would still have access to their accounts, whilst disallowing access to others without the required number of Picosiblings and the Pico device itself. The issue, here, appears to be what the user is expected to do to maintain the security of Pico, and if they are no longer in possession of their Pico or some number of their Picosiblings. 

%Key points of divergence in these perceptions arose from each participant's level of security consciousness and the presumed consequences of not protecting particular (important versus less important) accounts, such as those that involved the participant's own money.

%This involved evaluating the likelihood of the scheme failing due to interruptions to Pico reliability or security, causing both anxiety and annoyance. Annoyance and anxiety were also associated with the risk of losing or forgetting Picosiblings, even though the \textit{K out of N} scheme should mean that users may still access their accounts, whilst disallowing access to someone who does not have the required \textit{K out of N} Picosiblings and the Pico device itself.  

%In particular, the perceived risk of interruptions to Pico reliability and security was associated with annoyance and anxiety.  and the effort associated with having to prevent or deal with this

%Participants were more comfortable with the idea of carrying a dedicated Pico device if they already had experience of carrying a security token. In these cases, participants already took more responsibility for the security of their accounts and services, and Pico was not a big leap from that. This implies that 

In particular, the reluctance to accept Pico appears to lie with concerns about
taking on more personal \textbf{Responsibility} for authentication and recovery
-- the third and final identified phenomenon. Physical tokens are a more
tangible means of authentication than passwords.  Their use involves more
personal responsibility because the user must actively take on more inconvenient
strategies to maintain security and availability (e.g. due to losing or
forgetting the token).  The perception that responsibility for secure
authentication and recovery was being shifted towards the user seemed to amplify
participants' focus on the imagined risks of Pico, affecting their anxiety about
security and reliability, and anticipated annoyance with having to prevent or
deal with interruptions. For example:

\begin{quotation}
\noindent
``The worry would be obviously if you lost one and then you went to your access point and then realised that you lost one: \textbf{where would you always keep the spares?}... You wouldn't want to carry too many things" (Participant 17)
\end{quotation}

2FA schemes also increase personal responsibility, and so it follows that participants who had experience of other token-based schemes were more accepting of the idea of carrying a dedicated Pico. 

In the eyes of participants, the idea of Picosiblings seemed to delegate yet \emph{another} aspect of responsibility to the user: participants were concerned not only with how their accounts were protected, but how their Picosiblings were protected from loss and theft. The main suggested way of mitigating this anxiety was to hide or disguise Picosiblings. For example:

\begin{quotation}
\noindent
``This would be, like, \textbf{disguised as a key}. So, maybe, again this covert but then again also logical addition to something I already own" (Participant 16)
\end{quotation}

The interplay between \textbf{Inconvenience}, \textbf{Risk Perception}, and \textbf{Responsibility} is depicted in our final GT, in which we describe the process of \emph{acceptance of more tangible security}.

%which follows the discussion of these three factors in terms of passwords (something you know), an authentication token (something you have), and token-unlocking devices (other things you have). consisted of three parts, and was considered in terms of

\subsection{The Final Grounded Theory: Acceptance of More Tangible Security}
The problem of protecting more traditional forms of property, such as wallets and keys, has long-established solutions that people tend to understand better than protecting digital assets. Given that users already cope with carrying and managing physical items, one might expect users to be more accepting of the idea of carrying physical devices for authentication. However, our final GT suggests otherwise: although physical keys have stood the test of time, token-based authentication for computer systems seem to fall short of user requirements. 

\begin{figure*}
\centering
\includegraphics[width=0.475\textwidth]{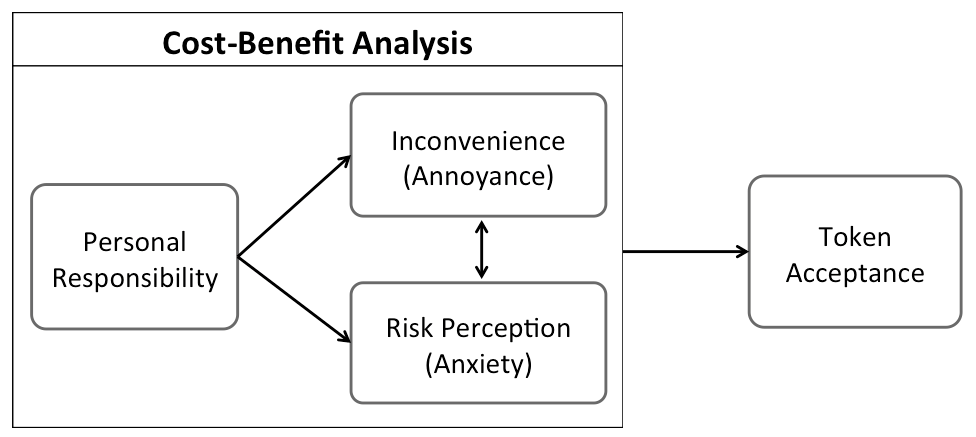}
\caption{The process involved in accepting a token-based authentication scheme (the central phenomenon) based on inconvenience, risk perception, and personal responsibility (core categories).}
\label{FinalGT}
\end{figure*}

Much of the discomfort with a token-based scheme can be traced back to participants' mental models of risk and the inconvenience associated with mitigating risk. This risk appears greater when the user considers physical devices, which are more tangible and thus considered ``easier to steal". After all, loss and theft of tangible property is a longer existing problem in human history than threats associated with much faster-paced technology advances. Thus, protecting physical devices is a more salient and intuitive problem than protecting digital information (passwords). The storyline below, used to present our final GT, describes how physical devices make users more overtly responsible for their own secure authentication, and creates an uncomfortable tension between a desire for day-to-day convenience and for avoiding the risk of loss and theft. We provide an illustration depicting this process (Figure \ref{FinalGT}).

The process of \emph{Token Scheme Acceptance} depicted in Figure \ref{FinalGT} positions the inconvenience of strategies for dealing with a token-based scheme as a precursor for whether it is accepted. This interacts with the user's risk perception, also affecting scheme acceptance. Potential users become more sensitive to inconvenience and risk when the level of perceived personal responsibility for authentication is high, as with managing physical tokens. This is because physical tokens, which are more tangible than knowledge-based secrets, literally put security in the hands of the user: with passwords, users passively place trust in the \emph{service} (say, an email provider) to reliably and securely authenticate them; a token requires trust in a physical device carried \emph{by the user} to reliably and securely authenticate them, making failure to login, or to keep authentication secure, something that the user has to go out of their way to deal with. This involves taking a much less convenient route than the current means of recovery by resetting one's password. 

More devices (Picosiblings) translates into even greater perceived likelihood of loss and theft. Even though a system of Picosiblings might have dispelled concerns about losing the \emph{Pico}, it increased concerns about losing Picosiblings. These scenarios were not only sources of interruptions (inconvenience), causing annoyance, but possible security breaches (risk), causing anxiety. 

This makes potential users of a token-based scheme, such as Pico, uncomfortable because it represents a significant behavior change that is not generally deemed worthwhile, evidenced in comments about Pico having to sufficiently and consistently improve lives before it would be accepted. This calls for improvements in the way Pico can be managed, and how we communicate Pico to potential users so that they are able to form an appropriate mental model.

\section{Discussion}
Our GT depicts token scheme acceptance as the result of evaluations of inconvenience and risk, amplified by the locus of responsibility for physical devices. We take the first steps towards validating our GT by comparing its processes with existing processes reported in the literature. This review of related work was left to the end so we could be confident that our final theory and model emerged from the data. 

%To deal with this, participants focused on device-management strategies, reflected in comments about protecting ``important" accounts with more Picosiblings, and about keeping these hidden. 

%Without this, we rely on the strategies that potential users conceive to deal with anticipated inconveniences and risks, if the system is not rejected outright.

%Loss and theft were particularly salient issues for participants. Our token is potentially safeguarded against theft, more so than passwords, because it relies on the presence of token-unlocking devices. Few participants recognized the benefit of this scheme, and instead focused on the effort associated with mitigating the risk of loss and theft, influencing the perceived inconvenience of our token-based scheme relative to current strategies for managing passwords. The take home message from participants appeared to be: \emph{``If you're going to make me more responsible for authentication, it had better be easy for me to fulfill my responsibilities"}.  

%and instead focused on the inconvenience and associated risks of losing these additional devices or having them stolen. 

\subsection{Related Research}
Our data consistently revealed an emphasis on reducing inconvenience, which was related to minimizing cognitive and physical effort, identified in prior research \cite{beautement2009compliance}. Also in line with other research, participants constructed their own Pico-related threat models, just as users construct their own password-related threat models \cite{adams1999users}. In particular, they seemed to believe that loss of any device represented a compromise to their accounts. As such, Pico represented a significant and risky behavior change for participants, even though, from a security standpoint, a targeted attack with Pico is less likely than with passwords because it cannot be conducted remotely. Theft is even less likely to result in a security violation with Pico because it would be harder to obtain the required \emph{K out of N} Picosiblings to carry out such an attack.

One observation made by Adams and Sasse \cite{adams1999users} was that users evaluate security threats and the importance of security measures based on what they actually see, or in the case of passwords, do not see. The issue here is tangibility, since users rarely have direct experience of the consequences of a security breach. We suggest that physical devices increase the tangibility of security, encouraging users to take more personal responsibility for a token rather than abdicating responsibility to services that rely on passwords. This shifts the liability for loss and theft to the user, increasing both anxiety about risk, and anticipated annoyance associated with inconvenience. 

In the field of human-computer interaction, models have been developed for predicting technology adoption (e.g. Davis \textit{et al.} \cite{davis1989user}, Meuter \textit{et al.} \cite{meuter2005choosing}, Venkatesh \textit{et al.} \cite{venkatesh2003user}). These present several important antecedent predictors of initial technology use, including (but not limited to) inertia, technology anxiety, experience, and novelty-seeking \cite{meuter2005choosing}. Our results are in line with this: participants mostly accepted the current state of authentication with passwords (inertia), expressed concern about adopting Pico (anxiety), and were more willing to adopt Pico if they were familiar with the concept of security tokens (experience) or if they were motivated by the potential enjoyment of choosing and modifying items (novelty-seeking). Such models tend to focus on \textit{first-time} technology use: a bad first impression is hard to recover from and so emphasis is placed on encouraging technology \emph{adoption}. Our research, too, considers initial reactions to the concept of Pico, and might be applied in future research to other technology-adoption models. For example, the \textit{Inconvenience} and \textit{Risk Perception} core categories in our final GT might be likened to: the ``Efficiency" and ``Effectiveness" sub-constructs of usability defined by the ISO 9241-11 \cite{ISO}; the ``Ease-of-Use" and ``Usefulness" precursors to system use in the Technology Adoption Model (TAM)\cite{davis1993user}; or the ``Effort Expectancy" and ``Performance Expectancy" precursors of system use in the Unified Theory of Acceptance and Usage of Technology (UTAUT) \cite{aguirre2010gender}. 

As well as technology acceptance research, our results are consistent with findings relating specifically to physical tokens (2FA and password managers). 
%De Cristofaro \textit{et al.}\ \cite{de2013comparative}, for example, found that cognitive effort, ease-of-use, and trustworthiness influenced the likelihood of using token-based two-factor authentication rather than smartphone-accessible two-factor authentication. These were all factors emerging in our research, with reluctance to adopt a token-based scheme based on the perceived effort, lack of ease-of-use, and lack of trustworthiness. In particular, 
Similar to De Cristofaro \textit{et al.}'s \cite{de2013comparative} results, for example, users were more amenable to the idea of using Pico for work or banking. We suggest that this was because token-based authentication, like 2FA, is more effortful, and deemed more worthwhile if required by the context, or if the user already has experience of a similar scheme. Our results are also consistent with Krol \textit{et al.} \cite{krol2015they}, whose participants worried about theft, and preferred the idea of biometric authentication. Participants also disliked the physical effort associated with having to interact with Pico. This is in line with Krol \textit{et al.} \cite{krol2015they}, whose participants disliked the number of steps involved in using a token and wanted the authentication process to be automated. It is also in line with Weir \textit{et al.} \cite{weir2009user}, who found that of three online banking tokens, participants preferred the token that required the least amount of physical effort (a push-button token), followed by a card-activated token, and then the chip-and-PIN method.

%Our grounded theory suggests that anxiety about security is particularly high with the idea of using our token-based scheme because we are making users overtly aware of their ownership, not only of secure authentication to \emph{services} via a token, but of token-unlocking \emph{devices}. Also influencing the likelihood of acceptance were requirements for particular benefits that participants deemed important precursors to using our token-based scheme, such as there being a sufficient number of providers and a significant improvement to the authentication process.

%Thus far, we have considered the results of other studies using specific groups of people, such as university staff and students (e.g. \cite{stobert2014password}), Mechanical Turk users (e.g. \cite{de2013comparative}), banking customers (e.g. \cite{weir2009user}), and organization employees (e.g. \cite{adams1999users}). We extend this research by purposefully recruiting a range of participants. Specifically, we interviewed two computer scientist students and eighteen other participants with a range of backgrounds and knowledge (Appendix \ref{ppts}). From these participants emerged an overall preference for a token that was integrated into a phone, with some leeway for carrying a dedicated device. 

\subsection{Limitations}
Our research is not without limitations, the most obvious being a lack of ecological validity and a potential for acquiescence bias. Participants were aware we wanted to replace passwords, and there was likely some motivation to give ``desirable" answers. It should be noted, however, that we were careful to communicate, both verbally and in the informed consent form, that we were looking for ways of \emph{improving} Pico, which was in an early design-stage; the aim was to encourage honest opinions, even if negative. Indeed, even with possible social pressure evoked by an interview setting, participants were neither wholly complimentary about Pico, nor wholly derogatory about passwords.

It is important to note that we do not make claims about Pico's long-term adoption. The focus of this paper is on predicting the factors that might influence willingness to make a behavior change (to use Pico instead of passwords), making users co-producers in their own secure authentication. This forms the basis of ideas about what we might be able to do, in terms of future research, to improve this. 

\subsection{Future Research}

This research is part of a larger iterative design process of prototyping. Carrying out such research from an early design stage provides valuable feedback that can be used to form hypotheses about how users might find using Pico, to create newer versions of prototypes, and to reduce the risk of major changes becoming necessary later. For example, future research could test the acceptability and reliability of biometrics for unlocking Pico, and determine whether this really does reduce user anxiety, before it is actually implemented in working devices. Future research might also investigate the possible disparity between what users say they would be willing to carry and what they would actually carry. 

%Based on insights gained from the GT, future research might compare feelings of
%personal responsibility evoked by more tangible authentication schemes, such as
%Pico (or even other token-based schemes, such as FIDO), with the more diffused
%responsibility associated with traditional login and recovery methods, which
%tend to be more abstract. As part of this, we might explore ways of increasing
%user understanding of token-based authentication, thereby reducing perceptions
%of risk and inconvenience, e.g., by highlighting the similarities between
%carrying a token and carrying keys or credit cards. 

In addition to changes to the Pico system, future research should consider the role of service providers and the regulatory environment in incentivizing and protecting users. For example, legislation or voluntary codes that prohibit service providers from shifting liability to their users in the event of a security incident may ameliorate anxieties about adopting token-based authentication schemes.

Finally, the significant concern expressed by participants about loss and theft suggests that more research on recovery would be valuable. As well as the day-to-day experience, it would be worth focusing on making the recovery procedure easier  so as to reduce the anticipated inconvenience of and risk associated with loss or theft.

\section{Conclusions}
We presented the GT results of an interview-based study looking at perceptions of Pico. Based on the final GT model, we suggest that a token-based scheme, such as Pico, makes authentication more tangible; this increases perceptions of personal responsibility for mitigating security risks and managing physical items, which is potentially inconvenient and anxiety-provoking for users. If this is the case, the challenge is in minimizing the friction between security requirements and user requirements, communicating the security benefits, and making sure these outweigh the anxiety costs. Specifically, although putting responsibility for security in the hands of the user is potentially good for protecting accounts and services, we face three key challenges. First, we need to increase willingness to take on more personal responsibility for security by reducing annoyance and anxiety. Second, we want to avoid also putting system failures (issues of reliability) in the hands of the user. Third, we need to find a way of aligning potential users' mental models of Pico with how it actually works to reduce the likelihood that the scheme will be rejected before it is tried.

This research lays the groundwork for a new way of considering token acceptance. The long-term aim is to maximize the usability of our scheme, but the key contribution of this paper is in providing insight into the factors that may prevent users from adopting a more tangible token-based scheme, using Pico as an example. Specifically, it provides insight into how people might perceive physicality when it comes to authentication. Our final GT and the role of personal (versus diffused) responsibility on token scheme acceptance is in need of further exploration.

\section*{Acknowledgments}

We are grateful to the European Research Council for funding this research through grant StG 307224 (Pico). We are also grateful to the following HCI students for their initial input into very early paper prototype designs of the Pico: T. Brouwer, K. Phatpanichot, R. Dorrity, G. Liang, J. Luo, and E. J. Kay-Coles. We extend gratitude to A. Hutchings who took time to review the write-up and to give detailed feedback.



\bibliographystyle{IEEEtranS}

%\appendix
%Appendix A
%\section*{Core Interview Questions}
%\label{questions}

%\section*{Token-Unlocking Devices}
%\label{TUDs}

%Duplicates are counted individually. For example, there were two hairbands, of which only one might be selected; if both were selected, this would be counted as two items chosen. Pairs, such as cufflinks, are counted as single items.

%\medskip
%\begin{tabular}{ l l }
%Glasses (small) & Glasses (large)\\
%Beltloop & Key fob\\
%Freestanding card & Card in ID holder\\
%Bracelet/anklet (x2) & Necklace\\
%Earrings (pair) & Ring (x2)\\
%Badge & Cufflinks (pair)\\
%Wrist watch & Watch (clip)\\
%Key & Key ring (standard)\\
%Key ring (heavy duty) & Earphone dust protector\\
%Hairband (x2) & Hairclip (x2)\\
%Plastic coins (x6) & Magnetic clips (x6)\\
%\end{tabular}

%\section*{Participant Occupation}
%\label{ppts}

%\end{document}

\end{document}